\documentstyle[12pt,epsf]{article}
\title{{\normalsize{{\hskip 8.5cm} BIHEP-TH-2000-2}} \\ 
   Possibility of a Light Pulse with Speed Greater
than c }
\author{Xian-jian Zhou  \\
        Institute of High Energy Physics, Academia Sinica \\
        P.O.Box 918-4, Beijing 100039 \\
        People's Republic of China}
\date{}
\begin{document}
\maketitle

\begin{abstract}
In two models it is shown that a light pulse propagates from a vacuum into
certain media
with velocity greater than that of a light in a vacuum (c). By numerical
calculation the propagating properties of such a light are given.
\end{abstract}

\vspace{1.5cm}

Recently L. J. Wang and his collaborators in their experiment [1] have
found that the group velocity of a laser pulse in particularly prepared
atomic caesium gas can much exceed that of light in a vacuum (c). Here 
we will explore the problem from a theoretical piont of view. Many years
ago A. Sommerfeld rigorously proved [2] that the velocity of light pulse
can not exceed c in absorption media. We call this conclusion as basic
theorem afterwardz. In the following we will first examine why usually the
basic theorem holds and in what condition it will be violated. Then two
models are proposed, where with precise numerical calculation we will show
that the basic theorem is indeed violated. Therefore the properties of light
propagating in media of the models are given. We believe that these properties
may appears in more realistic models.

Let a light pulse propagating along x axis toward its positive direction in a
vacuum $(x < 0)$, at time t = 0 arriving at x = 0, and then entering into a
medium $(x \geq 0)$ afterwards. At x = 0, the amplitude of the light pulse
changes with time t as
\begin{eqnarray}
f(t) = \left \{ \begin{array}{cc}
F(t) &~~t \geq 0 \\
0    &~~t < 0 ~. \end{array} \right.
\end{eqnarray}
Usually f(t) can be rigorously expressed in Fourier integration as
\begin{eqnarray}
f(t) = Re \int^{\infty}_{0} A(n) e^{-int} dn~~.
\end{eqnarray}
For simplicity suppose there is no reflection of light at x = 0. The amplitude
of the light pulse entering into the medium $(x \geq 0)$ is [2]
\begin{eqnarray}
f(t,x) = Re \int^{\infty}_{0} A(n) e^{-int + ikx} dn~~,
\end{eqnarray}
where $k = n \mu(n)/c$ and $\mu(n)$ is the complex refractive index of the
medium, which depends on the frequency of incident light (dispersion). In
vacuum $(x < 0)$, the amplitude of the light pulse is
\begin{eqnarray}
g(t,x) &=& Re \int^{\infty}_{0} A(n) e^{-int + inx/c} dn~~~~~~~~~~(x < 0)
\nonumber \\
 &=& f(t - x/c ) = \left \{ \begin{array}{cc}
F(t - x/c) &t - x/c \geq 0 \\
0    &t - x/c < 0 ~. \end{array} \right.
\end{eqnarray}
The shape of the light pulse propagating in a vacuum does not change because
all its Fourier components in (4) propagate with the same velocity c and do not
decay. In particular, all these components completely cancel each other in the
space-time region $t - x/c < 0$ (or $\theta = t ~c/x < 1)$ as long as the
light pulse propagates in a vacuum. One may think that such a cancellation may
not occur when light pulse propagates in a medium because of dispersion and absorption.
But Sommerfeld proved that such cancellation also occurs when light pulse
propagates in absorption media, which is just the basic theorem. In his
proving Sommerfeld let
\begin{eqnarray}
f(t) = \left \{ \begin{array}{cc}
sin ~\nu ~t &t \geq 0 \\
0    &t < 0 ~, \end{array} \right.
\end{eqnarray}
so that (3) becomes [2]
\begin{eqnarray}
f(t,x) = \frac{1}{2\pi} ~ Re \int_{c} e^{-i(nt - n \mu(n)x/c)}_{
~~~~~~~~~~~~~/(n -\nu)} ~dn~~,
\end{eqnarray}
where $\mu(n)$ is taken as
\begin{eqnarray}
\mu^{2} = 1 + \frac{a^{2}}{n^{2}_{0} - 2i\rho n - n^{2}} ~~,
\end{eqnarray}
which is the Lorentz-Lorenz refraction formula. $a^{2}, n_{0}, \rho$ are the
constants of the medium. $n_{0}, \rho$ represent the characteristic absorption
frequency, damping constant of the medium. Usually $\rho > 0$, light propagating
in the medium decays. The integration path c in (6) is shown in Fig.1, which
is along the real axis of n from $+ \infty$ to $-\infty$ through $n = \nu$ by a
small semicircle in the upper half of the complex plane. $\mu(n)$ in (7) has
branch pionts:
\begin{eqnarray*}
U_{1,2} &=& -i\rho \pm \sqrt{n^{2}_{0} - \rho^{2}} ~, ~where ~~\mu = \infty ~~;
\nonumber \\
N_{1,2} &=& -i\rho \pm \sqrt{n^{2}_{0} + a^{2} - \rho^{2}} ~, ~where ~~\mu = 0 ~~.
\end{eqnarray*}
We joing $U_{1}$ to $N_{1}$ and $U_{2}$ to $N_{2}$ by two branch lines, which
lie in the lower half of the complex plane, when $\rho > 0$. Because there is
no singularity and branch lines of the integrand of (6) in the upper half plane,
one can replace the integration path c by u in Fig.1, which is parallel to the
real axis in the upper half plane. When u moves to infinity in the upper half plane,
$\mu \rightarrow 1$ and when $t - x/c < 0$, then (6) = 0, which is just the basic
theorem. But when $\rho < 0$ (this is our model 1), the branch lines $U_{1} N_{1}$
and $U_{2} N_{2}$ lie in the upper half plane and the integration path u is not
equivalent to the path c in (6). Now the equivalent path u should be taken as
$u_{1} + u_{2} + u_{3} + u_{4}$ in Fig.1. By the same argument above, integration
along $u_{1}$ vanishs. Integrations along a pair of $u_{2}$ cancel each other.
The remaining integations along $u_{3}$ and $u_{4}$ usually do not vanish because
the branch lines lie in them. Therefore when $\rho < 0$ (model 1), the basic
theorem does not hold. In the following we will do numerical integration of (6)
to show that the basic theorem is indeed violated and to see what happens. 
$\rho < 0$ means propagation of light in the medium is gain-assisted light
propagation.

From (6) we get
\begin{eqnarray}
f(t,x) = \frac{1}{2} ~ Re [ie^{\gamma \omega(\bar{\nu})} ] + \frac{1}{2\pi}
Re \int^{\infty}_{0} [e^{\gamma w (-z + \bar{\nu})} -e^{\gamma w(z+\bar{\nu})} ]
\frac{d z}{z} ~~,
\end{eqnarray}
where $\gamma w (\bar{n}) = -int + in\mu x/c; \gamma = x n_{0}/c,
\bar{n} = n/n_{0}, \bar{\nu} = \nu/n_{0}, \bar{a}^{2}
= a^{2}/n^{2}_{0}, \bar{\rho} = \rho/n_{0}$, all of them are
dimensionless. Let
\begin{eqnarray}
w(z) = X(z) + i Y(z) ~~.
\end{eqnarray}
When $z >> 1$,
\begin{eqnarray}
X(z) \sim -\bar{a}^{2} \bar{\rho}/z^{2} ~~, ~~~~~Y(z) \sim z(1-\theta) -
\bar{a}^{2}/(2z)~,
\end{eqnarray}
where $\theta = c t/x$.

In reference [2], the typical values of parameters are given as 

\begin{eqnarray}
n_{0} = 4 \times 10^{16} ~s^{-1}, ~~~~a^{2} = 1.24 ~~n^{2}_{0}, ~~~\rho =
0.07 ~n_{0} ~,
\end{eqnarray}
where the medium is solid or liquid. For gas $a^{2}$ is about 1.001.

The numerical integration for (8) is difficult when $\mid \gamma Y(z) \mid $
becomes very large and hence $e^{\gamma Y(z)}$ is a very fast oscillatory function of z.
In fact when $n_{0} = 4 \times 10^{16} ~s^{-1}$ and $x = 1cm, \gamma = \frac{4}{3} \times
10^{6}$ is very large. But if $\gamma < 100$ for example, i.e., $x < 7.5 \times
10^{-5}cm$, we can do numerical integration of (8) with high precision.
Now the integration of (8) is divided into
sum of $\int^{R}_{0}$ and $\int^{\infty}_{R}$. With $R \sim 100$ for example,
fast oscillatory integrand appears in $\int^{\infty}_{R}$, when $z >> 1$.
Using (10)
\begin{eqnarray*}
Re(e^{\gamma \omega (z)}) = e^{-\gamma \bar{a}^{2} \bar{\rho}/z^{2}} \left \{
cos[\gamma z (1-\theta)] cos \frac{\gamma \bar{a}^{2}}{2z} + sin[\gamma z(1-\theta)
] sin \frac{\gamma \bar{a}^{2}}{2z} \right \} ~~,
\end{eqnarray*}
where the fast oscillatory factors are seperated in forms of sin and cos functions.
Mathematica can do that kind of integration with high precision. As an example, let
$\bar{a}^{2} = 1.24, \bar{\rho} = 0.07, \bar{\nu} = 10, \gamma = 1, \theta = 0.98$
in (8), one may get $f(t,x) = -1.02057 \times 10^{-13}$, which should vanish exactly due to the
basic theorem $(\rho > 0, \theta < 1)$. Considering the amplidude of incident light
pulse is 1, $10^{-13}$ is a very high accuracy of calculation. Now let
$\bar{\rho} = -0.07$ and other parameters unchanged, $f(t,x)$ in (8) is 0.0630255,
which is a definite evidence to show that the basic theorem is violated indeed
when $\rho < 0$. Because $\gamma = xn_{0}/c, \theta = tc/x,
f(t,x)$ may be looked as a function of $\gamma$ and $\theta, h(\gamma, \theta)$. Let us
fix $\gamma = 1$, i.e., $x = 0.75 \times 10^{-6}cm$, and $\theta$ change from (-7)
to 4, i.e., the time t from $-7/n_{0}$ to $4/n_{0}$, the amplitude of light pulse
$h(1, \theta)$ in the medium is shown in Fig.2, where 
$\bar{a}^{2} = 1.24, \bar{\rho} = -0.07, \gamma = 1, \bar{\nu} = 10 (\nu = 10 ~n_{0})$.
Again when $\rho < 0, h(1, \theta)$ does not vanish and the basic theorem is violated.
In Fig.2 even if $t < 0$, the amplitude $h(1, \theta)$ still does not vanish, which
means that before the incident light pulse arrives at the medium, the light in the
medium is already produced. For convenience, we call the light produced in the
medium when $\theta < 1$ as fastlight and that when $\theta > 1$ as normal
light. When the basic theorem holds, the fastlight vansishes. Some maximal values
of light amplitude near $\theta = 1$ and their corresponding $\theta$ values are
listed in  table 1, where two characteristics are shown:

(1) The amplitude of fastlight is oscillatory and decays as $\theta \rightarrow
-\infty$;

(2) The period for each oscillation for $\theta$ less and near 1 are roughly around
$\theta = 6$, which means its frequency is near the characteristic frequency
$n_{0}$ of the medium (the corresponding period is $\theta = 2\pi)$. This near
equality is due to the fact that the Fourier components of the light pulse
having freqencies equal to or near $n_{0}$ are most gain-assisted when $\rho < 0$.
When $\theta > 1$, the normal light soon oscillates with the frequency $\nu$
of  the incident light pulse (corresponding period $\theta = 0.2 \pi$) and its
amplitude is a little bit lager than 1(1.00089). When $\bar{\nu} = 1$ with
other parameters $\bar{a}^{2} = 1.24, \bar{\rho} = -0.07, \gamma = 1$ unchanged the two
properties (1) and (2) remain, but the amplitude of the fastlight increases to 6.14
near $\theta = 1$. The amplitude of normal light inceases to 7.31 with frequency
$\nu = n_{0}$.

Model 2: the  $\rho$ in (7) depends on frequency n as
\begin{eqnarray}
\rho (n) = \left \{ \begin{array}{ll}
\rho_{1}  &\ell - b < n < \ell + b \\
\rho_{2}  &n \leq \ell - b ~~or~~ n \geq \ell + b ~~, 
\end{array}  \right.
\end{eqnarray}
which is not continuous and therefore $\mu(n)$ and the integrand in (6) are not
the analystic functions of n. We still can use numerical calculation to see
whether the basic theorem holds:

(a) $\bar{a}^{2} = 1.24, \rho_{1} = -0.07, \rho_{2} = 0.07, \ell = 1, b = 0.01,
\bar{\nu} =1, \gamma = 1$. Now light amplitude in the medium are gain-assisted when
$(1-b)n_{0} < n < (1+b)n_{0}$ and decays otherwise. Fastlight with above two properties
appears again and its amplitude near $\theta = 1$ is 3.81, while that of normal
light is about 4.5;

(b) $\rho_{1} = 0.02$ with other parameters unchanged as in
(a). Now light in the whole frequency range in the medium decays. Still
fastlight remains with the two properties, but its amplitude becomes
small (0.0757) near $\theta = 1$.

So we may conclude that if $\mu(n)$ in (6) is an analystic
function of n with singularities (such as poles or branch lines) appearing in  the
upper half plane of n, or $\mu(n)$ is not an analystic function of n at all,
the basic theorem in general may not hold and fastlight appears. The Fourier
components of a light pulse now will not cancel each other in the medium in the
space-time region $t - x/c < 0$. This is why fastlight appears.

How to measure the velocity of the light pulse when fastlight appears in the
medium? Suppose a light pulse produced in a source at $t = 0$, propagating a
distance $\ell_{1}$ in a vacuum, then going through a medium with thickness
$\ell_{2}$. Just after the medium a light pulse detector is put. Usually the
fastlight appears in the medium after the light pulse is produced and the 
amplitude of the fastlight is inceasing when the light pulse
is approaching to the medium. If the amplitude of fastlight is able to become
large enough to trigger the detector at $t = t_{1}$, one may take $v = (\ell_{1}
+ \ell_{2})/t_{1}$ as the velocity of the light pulse propagating from the source
to the detector, which certainly exceeds c.

Although the models proposed above are not completely realistic, we believe that
production of the fastlight and some its properties in the models may remain in
a more realistic model, which is under investigation now.

We would like to thank professors Gu Yi-fan and Dong Fang-xiao for their beneficial
discussion.

\vspace{0.4cm}

{\bf Table 1. Some maximal values of light amplitude $h(1,\theta)$ near
$\theta =1$ $~~~~~~~~~~~(\bar{a}^{2} = 1.24, \bar{\rho}=-0.07, \gamma = 1$ and $\bar{\nu}
=10$ in model 1)}

\vspace{0.4cm}

\begin{tabular}{|c|c|c|c|c|c|c|c|} \hline
$\theta$          &-46.86  &-40.70 &-34.57 &-28.44  &-22.35 &-16.29 &-10.27 \\ \hline
$h(1,\theta)$ &0.00909 &0.0125 &0.0172 &0.0232 &0.0316 &0.0413 &0.0580 \\ \hline 
\multicolumn{8}{c}{~} \\ \cline{1-6}
-4.59  &0.8714 &1.1508 &1.77912 &2.40744 &3.03575 &\multicolumn{2}{|c}{~} \\ \cline{1-6}
0.0742 &0.0635 &1.0089 &1.0089 &1.0089 &1.0089    &\multicolumn{2}{|c}{~} \\ \cline{1-6}
\end{tabular}





\begin{figure}[thb]
\centerline{\epsfysize 2.2 truein \epsfbox{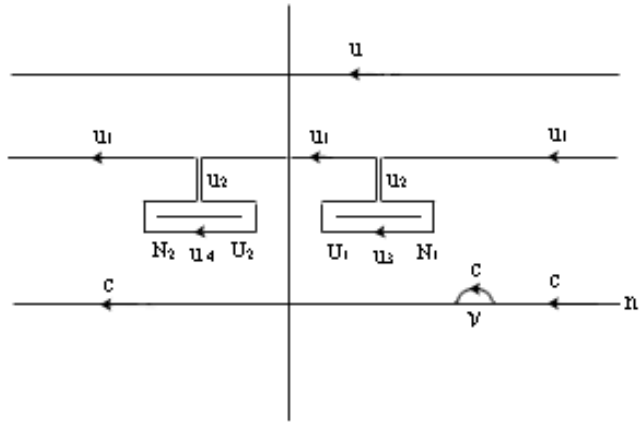}} 
\caption[]{The integration paths of (6).
}
\end{figure}

\begin{figure}[thb]
\centerline{\epsfysize 2.2 truein \epsfbox{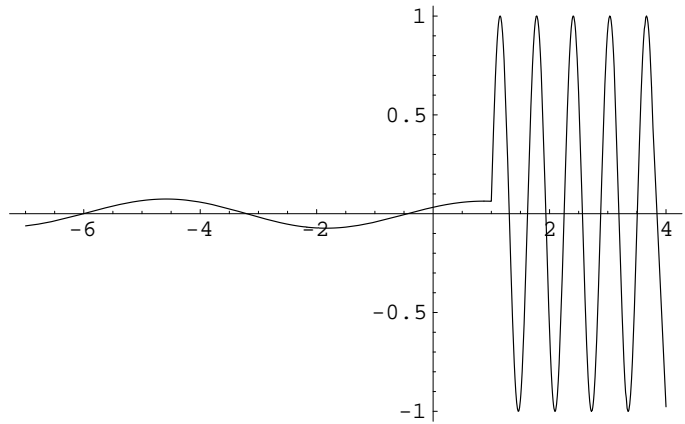}} 
\caption[]{Figure 2. $h(1,\theta)$ ($\bar{a}^{2} = 1.24, \bar{\rho} = -0.07, \gamma = 1$
and $\bar{\nu} = 10$ in model 1.)
}
\end{figure}

\end{document}